\documentclass[final,11pt]{elsarticle}
\usepackage[english]{babel}

\usepackage[utf8]{inputenc}
\usepackage{natbib}
\usepackage{lineno,hyperref}
\usepackage{lscape}
\usepackage{xcolor}
\usepackage{color}
\usepackage{graphics}
\usepackage{graphicx} 
\usepackage{colortbl}
\usepackage{setspace}
\usepackage{amssymb}
\usepackage{amsmath}
\usepackage{calrsfs}
\usepackage{multirow}
\usepackage{dsfont}
\usepackage{stmaryrd}
\usepackage{float}
\usepackage{lmodern} 
\usepackage{algorithm}
\usepackage{algpseudocode}
\usepackage[commentColor=cyan]{algpseudocodex}
\newcommand{\dr}{\textcolor{red}}
\DeclareMathOperator*{\argmin}{arg\,min}
\DeclareMathAlphabet{\pazccal}{OMS}{zplm}{m}{n}

\newcommand{\Var}{\mathrm{Var}}
\newcommand{\Cov}{\mathrm{Cov}}
\newcommand{\E}{\mathbb{E}}
\usepackage{enumitem}

\modulolinenumbers[5]
\usepackage{dcolumn}
\usepackage{bm}

\makeatletter
\def\ps@pprintTitle{%
  \let\@oddhead\@empty
  \let\@evenhead\@empty
  \let\@oddfoot\@empty
  \let\@evenfoot\@empty}
\makeatother

\usepackage{amsthm}
\usepackage[hang,flushmargin]{footmisc} 
\usepackage{graphics}
\usepackage{graphicx} 
\usepackage{mathrsfs}
\usepackage{bbm}
\usepackage{dsfont}
\usepackage{lmodern}
\usepackage{enumitem} 
\usepackage{subcaption}
\usepackage{booktabs}
\usepackage{sidecap}
\usepackage{xparse}
\usepackage{multirow}
\usepackage[margin=3cm]{geometry} 
\theoremstyle{definition}
\newtheorem{defn}{Definition}[section] 
\newtheorem{remark}{Remark}[section]
\newtheorem{prop}{Proposition}[section]
\newtheorem{lemma}{Lemma}[section]

\setcitestyle{square}

\DeclareFontFamily{U}{cbgreek}{}
\DeclareFontShape{U}{cbgreek}{m}{n}{
        <-6>    grmn0500
        <6-7>   grmn0600
        <7-8>   grmn0700
        <8-9>   grmn0800
        <9-10>  grmn0900
        <10-12> grmn1000
        <12-17> grmn1200
        <17->   grmn1728
      }{}
\DeclareFontShape{U}{cbgreek}{bx}{n}{
        <-6>    grxn0500
        <6-7>   grxn0600
        <7-8>   grxn0700
        <8-9>   grxn0800
        <9-10>  grxn0900
        <10-12> grxn1000
        <12-17> grxn1200
        <17->   grxn1728
      }{}

\DeclareRobustCommand{\qoppa}{%
  \text{\usefont{U}{cbgreek}{\normalorbold}{n}\symbol{19}}%
}

\makeatletter
\newcommand{\normalorbold}{%
  \ifnum\pdf@strcmp{\math@version}{bold}=\z@ bx\else m\fi
}

\begin{document}
\title{Randomized Kolmogorov–Smirnov Analysis of Volatility Roughness}

\author[inst1]{Sergio Bianchi}

\affiliation[inst1]{organization={MEMOTEF, Sapienza University of Rome},Department and Organization
            country={Italy}}

\author[inst1]{Daniele Angelini}


\begin{abstract}
We introduce a novel distribution-based estimator for the Hurst parameter of log-volatility, leveraging the Kolmogorov–Smirnov statistic to assess the scaling behavior of entire distributions rather than individual moments. To address the temporal dependence of financial volatility, we propose a random permutation procedure that effectively removes serial correlation while preserving marginal distributions, enabling the application of the KS framework to dependent data. We establish the asymptotic variance of the estimator, useful for inference and confidence interval construction. From a computational standpoint, we show that derivative-free optimization methods, particularly Brent’s method and the Nelder–Mead simplex, achieve substantial efficiency gains relative to grid search while maintaining estimation accuracy. Empirical analysis of the CBOE VIX index and the 5-minute realized volatility of the S\&P 500 reveals a statistically significant hierarchy of roughness, confirming previous findings that implied volatility is smoother than realized volatility. Both measures, however, exhibit Hurst exponents well below 1/2, reinforcing the rough volatility paradigm and highlighting the open challenge of disentangling local roughness from long-memory effects in fractional modeling.
\end{abstract}

            
\maketitle

\section{Introduction}
A large body of empirical work in financial econometrics has shown that measures of asset-price variability, whether based on returns or realized proxies, exhibit strong and persistent autocorrelation \citep{Ding1993, Baillie1996, Cont2001, Andersenetal2003, Baillieetal2019}. 
This stylized fact motivated the development of long-memory volatility models, i.e.\ processes with non-integrable autocorrelation functions \citep{PoonandGranger2003}. A canonical example is the Fractional Stochastic Volatility (FSV) model \citep{ComteRenault1998}, 
where the log-volatility follows a fractional Ornstein--Uhlenbeck (fOU) process driven by a fractional Brownian motion (fBm) with Hurst index $H>1/2$, thereby generating long-range dependence.

More recently, advances in option pricing and implied-volatility asymptotics \citep{Alòsetal2007, Fukasawa2019}, together with empirical evidence on realized-volatility dynamics \citep{GatheralJaissonRosenbaum2018}, have shifted attention to \emph{rough} volatility models. In particular, inspired by FSV, \cite{GatheralJaissonRosenbaum2018} introduce the Rough Fractional Stochastic Volatility (RFSV) model, replacing the FSV regime $H>1/2$ with $H<1/2$ to capture the pronounced high-frequency irregularity observed in data. Rough models, however, raise conceptual issues: due to the self-similarity of fBm, the Hurst parameter simultaneously governs path roughness and the asymptotic decay of correlations, implying that the fOU structure cannot disentangle roughness from long memory. As a consequence, for $H\in(0,1/2)$, both FSV and RFSV specifications produce short-range dependence combined with high local irregularity. \cite{Bennedsenetal2021} address this limitation by proposing a new class of continuous-time 
volatility models capable of accommodating both roughness and slowly decaying autocorrelations. The central role of the Hurst exponent in determining the appropriate modeling class has prompted the development of numerous estimation methods, including power spectral density estimation \citep{GatheralJaissonRosenbaum2018}, wavelet-based approaches \citep{abry2009wavelet}, quadratic-variation techniques \citep{rosenbaum2019estimation,SanchezGraneroetal2012,Trinidadetal2012}, and Bayesian or likelihood procedures \citep{bennedsen2021learning}, with recent extensions to deep-learning architectures \citep{bayer2021deep}.

Since realized volatility is only a discretized proxy for the latent instantaneous volatility process $\sigma_t$ and its integrated counterpart $\int \sigma_s\,ds$, estimation of the Hurst parameter based directly on it may be severely biased due to measurement error induced by discrete sampling. In particular, \citep{Fukasawaetal2022} show that regression-based estimators applied to daily realized volatility constructed from five-minute returns yield Hurst estimates close to $0.1$, largely independent of the true Hurst exponent of the underlying process. They attribute this effect to measurement error in realized volatility. Similarly, \citep{ContDas2022} demonstrate that roughness estimators based on pathwise variation of realized volatility systematically underestimate the Hurst parameter, implying that estimates below $1/2$ do not, in themselves, constitute evidence of intrinsic roughness of spot volatility.

To address this issue, several studies have proposed parametric estimators that explicitly model measurement error in realized volatility. Under the assumption that log prices follow a continuous semimartingale, \citep{Fukasawaetal2022} derive a limit theorem showing that the measurement error between realized and integrated volatility is asymptotically Gaussian with mean zero and variance $2/m$, where $m$ denotes the number of intraday observations per day, and construct a quasi-maximum likelihood estimator accordingly. Applied to five-minute realized volatility of major equity indices, this method produces Hurst estimates in the range $0.02$--$0.06$. Along similar lines, \citep{BolkoChristensenPakkanenVeliyeva2022} propose a GMM-based estimator under the same semimartingale framework and report comparably low Hurst estimates (below $0.05$), indicating extremely rough volatility dynamics.

To mitigate biases associated with moment-based estimators \citep{AngeliniBianchi2023}, we introduce a novel distribution-based estimator of the self-similarity parameter, $\hat{H}_0$, building on the framework of \cite{Bianchi2004}. Unlike moment-based estimators, our method employs the Kolmogorov--Smirnov (KS) statistic to compare entire distributions of appropriately rescaled increments, thus reducing nonlinear biases that may affect traditional approaches \citep{AngeliniBianchi2023}. Since the KS statistic is designed for independent samples, we incorporate a random-permutation device that eliminates serial dependence while preserving marginal distributions (Proposition~\ref{prop:commute}). Combined with a tailored optimization routine, this allows the KS methodology to be validly applied to dependent financial data. We further derive the estimator’s variance (Proposition~\ref{prop:varH}), enabling confidence intervals and clarifying its asymptotic behavior. This result is consistent with the Proposition 5 of \citep{bianchi2025roughness}, which was proved using Fisher's information.

Applying the estimator dynamically to both the VIX and 5-minute realized volatility of the S\&P~500 reveals that implied volatility is markedly smoother than realized volatility, in line with earlier findings \citep{Livieri02092018, floch2022}.  Nonetheless, both measures display Hurst exponents significantly below $1/2$, underscoring the relevance of the open problem of separating roughness from memory in fractional-volatility models.

The paper is organized as follows: section \ref{Method} recalls some preliminaries and describes the methodology; section \ref{Optimization} compares several optimization methods of the algorithm based on the method discussed in section \ref{Method}; section \ref{Pipeline} summarizes the entire pipeline estimation that will be used in the data analysis; section \ref{Implementation} provides and discusses the estimates, while conclusions are presented in section \ref{Conclusion}.

\section{Methodology} \label{Method}
\subsection{Preliminaries}
This section reviews the fundamental concepts underlying the Hurst parameter estimation method proposed by \citep{Bianchi2004}. Since these are largely established concepts, only the definitions will be provided, without delving into their respective justifications.

\begin{defn}
A nontrivial stochastic process $X=\{X_t\}_{t\geq0}$, stochastically continuous at $t=0$, is said to be $H_0$-self-similar with $H_0>0$ ($H_0$-ss) if for every $a \in \mathbb{R}^+$ and every finite set of times $t_1,\dots,t_k$,
\begin{equation}\label{eq:H-ss}
\bigl(X_{a t_1},\dots,X_{a t_k}\bigr) \stackrel{d}{=} \bigl(a^{H} X_{t_1},\dots,a^{H} X_{t_k}\bigr).
\end{equation}
\end{defn}

\begin{remark}\label{remark:0}
In general, a stochastic process is of course fully characterized by its finite-dimensional distributions, defined for any finite collection of times $t_1, \dots, t_n$ as:
\begin{equation*}
    F_{t_1,\ldots,t_n}(x_1,\ldots,x_n) = \mathbb{P}(X_{t_1}\leq x_1,\ldots,X_{t_n}\leq x_n).
\end{equation*}
Clearly, when observing a single path $\{\mathbf{x}_t\}_{t=1,\ldots,T}$ of a stochastic process $\{X_t\}$ -- which is the case covered by our work --, we are limited in what we can infer about its underlying distributions due to the lack of repeated, independent realizations. To estimate a joint distribution $F_{t_i, t_j}$ from data, many independent observations of the random vector $(X_{t_i}, X_{t_j})$ would be required, but a single path provides only one paired observation $(\mathbf{x}_{t_i}, \mathbf{x}_{t_j})$ for any given pair $(t_i, t_j)$. This is statistically insufficient to estimate the entire joint distribution and the best we can do is to construct the empirical marginal distribution. Under the assumption of stationarity, we treat the observations ${\mathbf{x}_1, \dots, \mathbf{x}_T}$ as draws from a common distribution and compute the empirical CDF $\hat{F}(x) = \frac{1}{T} \sum_{t=1}^T \mathbf{1}_{{\mathbf{x}_t \leq x}}$, where $\mathbf{1}$ is the indicator function. This limitation is actually the reason why the random permutation method in Section \ref{sec:RandomPermutation} allows to test a necessary condition of self-similarity, not self-similarity in itself. Despite this, for large classes of stochastic processes the condition suffices to infer the self-similarity of the finite-dimensional distributions and not only of the marginal ones (see Propositions \ref{prop:commute}, \ref{prop:Sigma-marginal-fBm} and \ref{prop:Levy_from_marginal}).
\end{remark}

\begin{remark}\label{remark:1}
    If a $H_0$-$ss$ stochastic process $\{X_t\}_{t\geq0}$, with $X_0=0$ a.s., has stationary increments, the increment process is self-similar with the same parameter ($H_0$-$sssi$):
    \begin{equation}\label{eq:H-sssi}
        \{X_{t+a}-X_t\}_{t\geq0} \overset{d}{=} \{a^{H_0}(X_{t+1}-X_t)\}_{t\geq0}, \,\,\, \forall a\in\mathbb{R}^+.
    \end{equation}
\end{remark}
Building upon the approach presented in \citep{Bianchi2004}, the self-similarity exponent $H_0$ can be estimated by testing the equality of the distributions in equation \eqref{eq:H-ss} (or equation \eqref{eq:H-sssi}). To accomplish this, the Kolmogorov-Smirnov test may be employed.\\

In fact, given a compact timescale set $\pazccal{A}=\left[\underline{a},\overline{a}\right]\subset\mathbb{R}^+$ and denoted by $\Phi_{X_{t}}(x)$  the cumulative distribution function of the stationary increment process $Z_{t,a}:=X_{t+a}-X_t$, for any $a\in\pazccal{A}$, equality  \eqref{eq:H-sssi} can be written as
\begin{equation}\label{eq:distr_H-sssi1} \nonumber
    \Phi_{Z_{t,a}}(x) := \mathbb{P}\left(Z_{t,a}<x\right) \overset{\text{using eq. \eqref{eq:H-sssi}}}{=} \mathbb{P}\left(a^{H_0}Z_{t,1}<x\right) = \Phi_{Z_{t,1}}\left(a^{-H_0}x\right),
\end{equation}
\begin{equation}\label{eq:distr_H-sssi2} \nonumber
    \Phi_{a^{-H}Z_{t,a}}(x) := \mathbb{P}\left(a^{-H}Z_{t,a}<x\right) \overset{\text{using eq. \eqref{eq:H-sssi}}}{=} \mathbb{P}\left(a^{H_0-H}Z_{t,1}<x\right) = \Phi_{Z_{t,1}}\left(a^{H-H_0}x\right),
\end{equation}
and
\begin{equation}\label{eq:distr_H-sssi3}
    \delta_{Z_{t}}\left(\Psi_H\right) = \sup\limits_{x\in\mathbb{R}}\left\vert \Phi_{Z_{t,1}}\left(\underline{a}^{H-H_0}x\right)-\Phi_{Z_{t,1}}\left(\overline{a}^{H-H_0}x\right)\right\vert,
\end{equation}
where $\Psi_H := \{\Phi_{a^{-H}Z_{t,a}}(x),a\in\pazccal{A},x\in\mathbb{R}\}$ is the set of the distribution functions of $\{a^{-H}Z_{t,a}\}_{t\geq0}$ and $\delta_{Z_t}$ is the diameter of $\Psi_H$ on the metric space $\left(\Psi_H,\qoppa\right)$, where $\qoppa$ is the distance function induced by the sup-norm $\vert\vert\cdot\vert\vert_{\infty}$ with respect to $\pazccal{A}$.
It is easy to recognize in \eqref{eq:distr_H-sssi3} the Kolmogorov-Smirnov statistic.\\

The diameter $\delta\left(\Psi_H\right)$ satisfies the following propositions (see \citep{Bianchi2004} for proofs):
\begin{prop}\label{prop1}
    $\{Z_{t,a}\}_{t\geq0}$ is $H_0$-$ss$ if and only if, for any $\pazccal{A}\subset\mathbb{R}^+$, $\delta_{Z_t}\left(\Psi_{H_0}\right)=0$.
\end{prop}
\begin{prop}\label{prop2}
    Let $\{Z_{t,a}\}_{t\geq0}$ be $H_0$-$ss$. Then $\delta_{Z_t}\left(\Psi_H\right)$ is non-increasing for $H\leq H_0$ and non-decreasing for $H\geq H_0$.
\end{prop}
\begin{prop}\label{prop3}
    Let $\{Z_{t,a}\}_{t\geq0}$, \textbf{x}$\geqq0$ or \textbf{x}$\leqq0$, $\{\pazccal{A}_i\}_{i=1,\ldots,n}$ be a sequence of timescale sets such that, denoted by $\underline{a}_i=\min\left(\pazccal{A}_i\right)$ and by $\overline{a}_i=\max\left(\pazccal{A}_i\right)$, it is $\underline{a}_i\leq\underline{a}_j$ and $\overline{a}_i\geq\overline{a}_j$ for $i>j$. Then, with respect to the sequence $\{\pazccal{A}_i\}$, $\delta_{Z_t}\left(\Psi_H\right)$ is non-decreasing for $H\neq H_0$ and zero for $H=H_0$.
\end{prop}

\begin{prop}\label{prop4}
Let $B^{H_0}_t$ be a fractional Brownian motion (fBm) with parameter $H_0$. Then
\begin{eqnarray} \label{eq:diamfbm}
    \delta_{B^{H_0}_t}^1(\Psi_H)&=&|\Phi(x\underline{a}^{H_0-H})-\Phi(x\overline{a}^{H_0-H})|\\
    &=& \int_{x\overline{a}^{H_0-H}}^{x\underline{a}^{H_0-H}}\varphi(u)du
\end{eqnarray}
where $\varphi(u)=\frac{1}{\sqrt{2\pi}}e^{-u^2/2}$ and $\Phi(z)=\int_{-\infty}^z \varphi(u)du$.
\end{prop}
Two aspects must be noted with regard to Proposition \ref{prop4}:
\begin{itemize}[leftmargin=*]
    \item in addition to the argument in Remark \ref{remark:0}, diameter \eqref{eq:diamfbm} cannot be extended to the full set of finite-dimensional distributions, since the loss of one-dimensional ordering and monotonicity, combined with the fact that the multivariate Kolmogorov-Smirnov statistic is neither distribution-free nor available in closed form, prevents such a reformulation. However, Proposition \ref{prop:Sigma-marginal-fBm} ensures that this is not a problem, because for the fBm the paramater $H_0$ estimated for the one-dimensional case allow to infer $H_0$ for the whole process;
    \item although derived for fBm, diameter \eqref{eq:diamfbm} is informative well beyond the fBm class, since it applies also to processes whose local behaviour is asymptotically equivalent to that of an fBm. This is particularly valuable in those cases, where a wide range of stochastic volatility models—most notably the fractional Ornstein–Uhlenbeck (fOU) process—exhibit local fBm–like structure. Other examples include multifractional Brownian motion, Volterra-type fractional diffusions, rough Bergomi and related rough-volatility models, as well as Multifractional Processes with Random Exponent (MPRE).
\end{itemize}

Denoted by $Z_t$ the log-volatility process and by $\Psi_H$ the set of rescaled distribution of its increments, the idea is to estimate the self-similarity parameter $H_0$ by seeking the value of $H\in\left(0,1\right]$ that minimizes the Kolmogorov-Smirnov statistic $\delta_{Z_{t}}\left(\Psi_H\right)$ of any pair of rescaled distributions of $Z_{t,a}$. In fact, Proposition \ref{prop1} ensures that if $Z_{t,a}$ is $H_0$-ss then $\delta_{Z_t}\left(\Psi_H\right)$ has a unique minimum with respect to $H\in\left(0,1\right]$, in correspondence of the value $H_0$. Thus,
\begin{equation}\label{eq:empirical_H}
    \hat{H}_0 = \argmin_{H\in\left(0,1\right]}\,\,\hat{\delta}_{Z_t}\left(\Psi_H\right)
\end{equation}
The methodology outlined above raises two main challenges:
\begin{itemize}[leftmargin=*]
    \item assessing the statistical significance of the minimum, since the Kolmogorov--Smirnov (KS) distribution is valid only under the assumption of independent and identically distributed samples. This requires modifying the input processes in \eqref{eq:distr_H-sssi3} to remove potential dependence and enforce conformity with the KS distribution. This can be achieved using results from random permutation theory, briefly reviewed below;
    \item quantifying the variance of the estimator \eqref{eq:empirical_H}, which is essential for constructing confidence intervals and evaluating deviations of the parameter—possibly time-varying—from the equilibrium level implied by the market efficiency hypothesis.
\end{itemize}

\subsection{Random permutations} \label{sec:RandomPermutation}
Since the power spectrum $S_{Z}(\omega)$ of a stationary process $\{Z_n\}_{n\in\mathbb{Z}}$ is defined as the Fourier transform of its covariance function, the autocovariance function can be obtained as the inverse Fourier transform of its power spectrum, i.e.
    \begin{equation}
        K_{Z}(q)=\mathbb{E}[Z_nZ_{n-q}] = \int_{-\pi}^{\pi}e^{iq\omega}S_Z(\omega)d\omega.
    \end{equation}
Applying this to the power spectrum of the fractional Gaussian noise $Z^H_{t,a}$, $a>0$, one has
    \begin{equation}\label{eq:powerfgn}
        S_{Z^H}(\omega) = 2 c_H(1-\cos\omega)\sum_{j\in\mathbb{Z}}\vert2\pi j + \omega\vert^{-1-2H}, \;\;\;\;\;\forall\omega\in[-\pi,\pi],
    \end{equation}
    with $c_H = \frac{C^2}{2\pi}\sin(\pi H)\Gamma(2H+1)$ \citep{Coeurjolly2000}.
The structure of \eqref{eq:powerfgn} can be disrupted by random permutations: given the random sequence $Z = \{Z_\ell, \ell\in\mathbb{Z}\}$, a new sequence $U = \{U_\ell, \ell\in\mathbb{Z}\}$ is defined such that 
 \begin{equation}\label{eq: new sequence}
     U_\ell = Z_{\overline{\ell}L+b_{\underline{\ell}}}, \;\; \ell = \overline{\ell}L+\underline{\ell},\;\; \overline{\ell}\in\mathbb{Z},\;\; 0\leq\underline{\ell}\leq L-1,
 \end{equation}
 where $b = (b_0, b_1, \ldots, b_{L-1})$ is a random permutation of the vector $(0,1,\ldots,L-1)$, uniformly distributed such that $\mathbb{P}(b) = \frac{1}{L!}$, with $\overline{\ell}$ and $\underline{\ell}$ the quotient and the remainder, respectively, of $\frac{\ell}{L}$.
Based on Section 4 in \citep{LacazeRoviras2002}, \citep{AngeliniBianchi2025} prove the following \\

 \begin{prop}\label{prop:1}
     Let $Z^H = \{Z^H_{\ell,a},\ell\in\mathbb{Z},a>0\}$ be a fractional Gaussian noise. Then, the new sequence $\tilde{Z}^H = \{Z^H_{\overline{\ell}L + b_{\underline{l}}+\phi,a},\ell\in\mathbb{Z},a>0\}$, with an independent random phase $\phi$ uniformly distributed on the set of integers $(0,1,\ldots,L-1)$, has factorizable covariance in the limit of large permutation length $L$
     \begin{equation}
         \lim\limits_{L\to\infty}K_{\tilde{Z}^H}(q) = \mathbb{E}\left[Z^H_{\overline{\ell}L + b_{\underline{l}}+\phi,a}\right]\mathbb{E}\left[Z^H_{\overline{\ell+q}L + b_{\underline{l+q}}+\phi,a}\right] = 0
     \end{equation}
     for $q\neq 0$.
 \end{prop}

\begin{remark}
    Given the stationarity of a fractional Gaussian noise (fGn), aside from a multiplicative parameter, we can state that $S_{Z^H}(\omega)$ is the power spectrum of its autocorrelation function; since random permutation acts only on the order of the elements of a sequence and not on their distribution, the structure of self-similarity is maintained, i.e.
    \begin{equation}\nonumber
        \{Z_{i,a}^{H_0}\} \overset{d}{=} \{a^{H_0}Z_{i,1}^{H_0}\} \;\;\;\Rightarrow\;\;\;\{Z_{\overline{i}L+b_{\underline{i}}+\phi,a}^{H_0}\} \overset{d}{=} \{a^{H_0}Z_{\overline{i}L+b_{\underline{i}}+\phi,1}^{H_0}\}.
    \end{equation}
\end{remark}

Proposition \ref{prop:1} asserts that, for sufficiently large values of $L$, the randomly permuted sequence $\tilde{Z}^H$ can be regarded as a white noise process. For an fGn this occurs starting from $L \sim 100$, as shown in \cite{AngeliniBianchi2025}. The rate at which the power spectrum $K_{\tilde{Z}^H}(q)$ converges to that of white noise depends on the regularity of the original power spectral density  $S_{Z^H}(\omega)$. For example, the randomly permuted versions of NRZ-type (Non-Return-to-Zero) and Biphase processes approximate white noise behavior as $L \geq 500$ \citep{LacazeRoviras2002}. 

\begin{remark}
Random permutation preserves marginal distributions ($k=1$ in equation \eqref{eq:H-ss}), but generally destroys temporal/joint structure across lags. In general, testing $H_0$ based only on the marginal distributions of a process (even if those marginals exhibit the correct scaling $t^{H_0}$) does not imply that $H_0$ is the self–similarity parameter of the whole process. To characterize self–similarity one needs information about the joint finite–dimensional distributions
or impose additional structural assumptions under which the covariance structure determines all finite–dimensional laws. In general, it holds the following 

\begin{prop}[necessity]\label{prop:commute}
Let $X=(X_1,\dots,X_N)$ be a finite sample and $\pi$ be a permutation of the index set $\{1,\dots,N\}$. Define $Y_i := X_{\pi(i)}$. If $Y_t=X_{\pi(t)}$ is $H_0$-ss for all $a \in \mathbb{R}^+$, then $\pi$ must satisfy the commutation relation
\begin{equation}\label{eq:commute}
\pi(a t) = a \pi(t),
\end{equation}
for all $t \geq 0$ and $a\in \mathbb{R}^+$.
\end{prop}
\begin{proof}
    Let $X=(X_1,\dots,X_N)$ be a finite sample and $\pi$ be a permutation of the index set $\{1,\dots,N\}$. Defining $Y_i := X_{\pi(i)}$, $H_0$-ss of $Y$ would entail
\begin{equation*}
    \{Y_{at_1},\ldots,Y_{at_k}\} \overset{d}{=} \{a^{H_0}Y_{t_1},\ldots,a^{H_0}Y_{t_k}\}, \quad \forall a>0, t_1,\ldots,t_k.
\end{equation*}
that is
\begin{equation} \label{eq:comss}
\{X_{\pi(at_1)},\ldots,X_{\pi(at_k)}\} \overset{d}{=} \{a^{H_0}X_{\pi(t_1)},\ldots,a^{H_0}X_{\pi(t_k)}\}.
\end{equation}
On the other side, by $H_0$-ss of $X$, for any choice of indices $s_j$ we have
\begin{equation}\label{eq:commss2}
    \{X_{as_1},\ldots,X_{as_k}\}\overset{d}{=} \{a^{H_0}X_{s_1},\ldots,a^{H_0}X_{s_k}\}.
\end{equation}
The right-hand side of \eqref{eq:comss} already matches the right-hand side of \eqref{eq:commss2} if we take $s_j = \pi(t_j)$. Thus, the only way to use \eqref{eq:commss2} to replace the left-hand side of \eqref{eq:comss} is to require that the permutation $\pi$ commutes with the scaling map $S_a:t\mapsto at$, i.e. at the level of indices
\begin{equation*}
    X_{\pi(at_j)} = X_{a\pi(t_j)}
\end{equation*}
that is
\begin{equation*}
\pi(a t) = a \pi(t), \qquad \forall t \geq 0.
\end{equation*}
If \eqref{eq:commute} holds for every $a$, one has
\begin{equation*}
    \{X_{\pi(at_1)},\ldots,X_{\pi(at_k)}\} = \{X_{a\pi(t_1)},\ldots, X_{a\pi(t_k)}\}\overset{d}{=} \{a^{H_0}X_{\pi(t_1)},\ldots,a^{H_0}X_{\pi(t_k)}\}
\end{equation*}
and $Y$ inherits self-similarity.\\
For a finite index set $\{1,\dots,N\}$ and an integer scaling $a$, the set of permutations satisfying \eqref{eq:commute} is extremely small compared to $N!$. For large $N$, the probability that a uniform random permutation satisfies the commutation property is essentially zero. In the continuous case the probability is literally zero. Thus a random permutation almost surely destroys the commutation relation and hence self-similarity when this is tested on finite-dimensional distributions with $k \geq 2$.
\end{proof}
\end{remark}

Given Proposition \ref{prop:commute}, the technique we propose constitutes in general a necessary condition for self-similarity, since any self-similar process must have marginals that scale consistently. However, if one imposes additional assumptions that make the finite–dimensional distributions determined (or strongly constrained) by the one–dimensional marginals, then the self–similarity parameter $H_0$ estimated with respect to the marginal distributions suffices to characterize all finite-dimensional distributions. Two relevant examples are the fBm-type Gaussian processes and the Lévy processes, both recognized models for financial applications.

\paragraph{1. fBm-type Gaussian processes}
Since fBm-type processes are centered Gaussian process with stationary increments and continuous paths completely determined in law by their covariance function, scaling of the one–dimensional marginals forces an appropriate scaling of the covariance and self–similarity follows. In fact, it holds the following

\begin{prop}[Identification of the self--similarity exponent from marginal distributions]
\label{prop:Sigma-marginal-fBm}
Let $X = \{X_t\}_{t \ge 0}$ be a fractional Brownian motion (fBm) with
self-similarity parameter $H_0 \in (0,1)$ and 
$\Sigma$ be an estimator that, given the marginal
distributions $\{\mathcal{L}(X_t)\}_{t>0}$, 
returns the corresponding exponent $H$; that is, by definition of $\Sigma$,
\begin{equation*}
    \Sigma\bigl(\{\mathcal{L}(X_t)\}_{t>0}\bigr) = H
  \quad\text{whenever}\quad
  X_t \sim \mathcal{N}(0,\sigma^2 t^{2H})\ \text{for all }t>0.
\end{equation*}
Then:
\begin{enumerate}[leftmargin=*, itemindent=0pt, labelindent=0pt]
  \item $\Sigma\bigl(\{\mathcal{L}(X_t)\}_{t>0}\bigr) = H_0$. 
  \item 
  for every $a>0$ and every $0 \le t_1 < \ldots < t_n$, $(X_{a t_1},\dots,X_{a t_n})
          \stackrel{d}{=}
          (a^{H_0} X_{t_1},\dots,a^{H_0} X_{t_n})$,
that is, estimating the self--similarity exponent on the marginal distributions is sufficient to infer the self--similarity exponent of the full fBm (i.e. of all its finite--dimensional distributions).
\end{enumerate}
\end{prop}

\begin{proof}
By the covariance representation of fBm, $\Cov(X_s,X_t) = \frac{1}{2}\Bigl(s^{2H_0} + t^{2H_0} - |t-s|^{2H_0}\Bigr)$ $(s,t \geq 0)$, it readily follows $\Var(X_t)=t^{2H_0}$, $t \geq 0$. Thus $X_t \sim \mathcal{N}\bigl(0,t^{2H_0}\bigr)$.
Comparing with $\Var(X_t) = \sigma^2 t^{2H}$, we have $\sigma^2 = 1$ and
$H = H_0$, hence by the defining property of $\Sigma$,
\[
  \Sigma\bigl(\{\mathcal{L}(X_t)\}_{t>0}\bigr) = H_0,
\]
which proves (1).

For (2), fix $a>0$ and define $Y_t := a^{-H_0} X_{a t}$, $t \geq 0$. Then 
\begin{align*}
  \Cov(Y_s,Y_t)
  &= a^{-2H_0} \Cov(X_{a s},X_{a t}) \\
  &= a^{-2H_0}\,\frac12\Bigl((a s)^{2H_0} + (a t)^{2H_0}
                             - |a t - a s|^{2H_0}\Bigr) \\
  &= \frac12\Bigl(s^{2H_0} + t^{2H_0} - |t-s|^{2H_0}\Bigr)
   = \Cov(X_s,X_t).
\end{align*}
Thus $X$ and $Y$ are centered Gaussian processes with the same covariance
function, hence the same finite--dimensional distributions. Therefore, for
every $0 \le t_1 < \dots < t_n$,
\begin{equation*}
  \{Y_{t_1},\dots,Y_{t_n}\} \stackrel{d}{=} \{X_{t_1},\dots,X_{t_n}\},
\end{equation*}
i.e.
\begin{equation*}
    \{a^{-H_0} X_{a t_1},\dots,a^{-H_0} X_{a t_n}\} \stackrel{d}{=}\{X_{t_1},\dots,X_{t_n}\},
\end{equation*}
which is equivalent to
\begin{equation*}
    \{X_{a t_1},\dots,X_{a t_n}\} \stackrel{d}{=} \{a^{H_0} X_{t_1},\dots,a^{H_0} X_{t_n}\}.
\end{equation*}
Hence $X$ is $H_0$-self similar, and the exponent recovered from the marginals coincides with the self-similarity exponent of the full process. Since fBm $X$ is stationary with $X=0$ a.s., the argument trivially holds for the increment process as well.
\end{proof}

\begin{remark}
    The argument in the above Proposition holds locally for those processes which asymptotically behave like an fBm. These include the stationary fractional Ornstein-Ulhenbeck (fOU) process, pivotal model for stochastic volatility. The process is defined as the unique almost surely continuous process that solves the following stochastic differential Langevin-like equation,
\begin{equation}\label{eq:LangevinEq}
    dY_t^H = -\lambda Y_t^Hdt + \eta dB_t^H, \quad \lambda,\eta\in\mathbb{R}^+,
\end{equation}
driven by an fBm $B_t^H$ of Hurst exponent $H\in(0,1)$. The restriction of the process to $t\geq 0$ is
\begin{equation}
    Y_t^H = \eta e^{-\lambda t}\int_{-\infty}^t e^{\lambda u}dB_u^H
\end{equation}
with the initial condition $Y_0^H = \eta\int_{-\infty}^0 e^{\lambda v}dB_v^H$ and zero long-mean term~\cite{CheriditoKawaguchiMaegima2003}. The coefficients $\lambda$ and $\eta$ represent, respectively, the mean-reversion and diffusion parameters.
\end{remark}

\paragraph{2. Lévy (independent increment) case} A second important example regards Lévy processes. The assumption of stationary and independent increments makes all finite–dimensional distributions determined by the one–dimensional distributions of the increments. In this case, it holds the following

\begin{prop} \label{prop:Levy_from_marginal}
Let $X = \{X_t\}_{t\ge 0}$ be a L\'evy process. Assume that for some $H_0>0$
\begin{equation*}
    X_t \;\overset{d}{=}\; t^{H_0} X_1,\qquad \forall\, t>0
\end{equation*}
(i.e.\ the marginals are $H_0$–self–similar). Then, for all $a>0$ and all
$0\le t_1<\dots<t_n$,
\begin{equation*}
    \{ X_{a t_1},\dots,X_{a t_n} \}
  \;\overset{d}{=}\;
  \{ a^{H_0} X_{t_1},\dots,a^{H_0} X_{t_n} \},
\end{equation*}
so the whole process is $H_0$–self–similar.
\end{prop}

\begin{proof}
For a L\'evy process, the one–dimensional characteristic function has the form
\begin{equation*}
    \phi_t(u) := \E\bigl[e^{i u X_t}\bigr] = e^{t\psi(u)},\qquad t\ge 0,\ u\in\mathbb{R},
\end{equation*}
for some characteristic exponent $\psi:\mathbb{R}\to\mathbb{C}$.\\

\noindent By the marginal self–similarity, for all $t>0$,
\begin{equation*}
    \E[e^{i u X_t}]
  = \E[e^{i u t^{H_0} X_1}]
  = \E[e^{i (t^{H_0} u) X_1}]
  = \phi_1(t^{H_0}u)
  = e^{\psi(t^{H_0}u)}.
\end{equation*}
Comparing with $\E[e^{i u X_t}] = e^{t\psi(u)}$ gives
\begin{equation*}
    \psi(t^{H_0}u) = t\,\psi(u),\qquad \forall\,t>0,\ u\in\mathbb{R}.
\end{equation*}
Setting $t=\lambda^{1/H_0}$ yields
\begin{equation*}
    \psi(\lambda u) = \lambda^{1/H_0}\,\psi(u),\qquad \forall\,\lambda>0,\ u\in\mathbb{R},
\end{equation*}
so $\psi$ is homogeneous of order $\alpha:=1/H_0$.\\

\noindent Next, recall the joint characteristic function of a L\'evy process:
for $0 \leq t_1<\dots<t_n$ and $u=(u_1,\dots,u_n)\in\mathbb{R}^n$, because of the stationary independent increments the classical formula holds (see e.g. \citep{Sato1999})
\begin{equation*}
    \E\Big[\exp\Big(i\sum_{k=1}^n u_k X_{t_k}\Big)\Big]
  = \exp\Bigg(
      \sum_{j=1}^n (t_j-t_{j-1})\,
      \psi\Big(\sum_{k=j}^n u_k\Big)
    \Bigg),
\end{equation*}
with $t_0=0$.\\

\noindent Apply this formula twice:\\

\noindent (\emph{i}) For $(X_{a t_1},\dots,X_{a t_n})$:
\begin{equation*}
    \E\Big[\exp\Big(i\sum_{k=1}^n u_k X_{a t_k}\Big)\Big]
  = \exp\Bigg(
      \sum_{j=1}^n a(t_j-t_{j-1})\,
      \psi\Big(\sum_{k=j}^n u_k\Big)
    \Bigg).
\end{equation*}

\noindent (\emph{ii}) For $(a^{H_0} X_{t_1},\dots,a^{H_0} X_{t_n})$:
\begin{equation*}
    \E\Big[\exp\Big(i\sum_{k=1}^n u_k a^{H_0} X_{t_k}\Big)\Big]
  = \exp\Bigg(
      \sum_{j=1}^n (t_j-t_{j-1})\,
      \psi\Big(a^{H_0}\sum_{k=j}^n u_k\Big)
    \Bigg).
\end{equation*}

\noindent Using homogeneity of $\psi$,
\begin{equation*}
    \psi\Big(a^{H_0}\sum_{k=j}^n u_k\Big)
  = a\,\psi\Big(\sum_{k=j}^n u_k\Big),
\end{equation*}

\noindent the exponent in (\emph{ii}) becomes
\[
  \sum_{j=1}^n (t_j-t_{j-1})\,\psi\Big(a^{H_0}\sum_{k=j}^n u_k\Big)
  = \sum_{j=1}^n a(t_j-t_{j-1})\,\psi\Big(\sum_{k=j}^n u_k\Big),
\]
which is exactly the exponent in (\emph{i}). Hence the two joint characteristic
functions coincide for all $u\in\mathbb{R}^n$, and therefore
\begin{equation*}
( X_{a t_1},\dots,X_{a t_n} )
  \overset{d}{=}
  ( a^{H_0} X_{t_1},\dots,a^{H_0} X_{t_n} ),    
\end{equation*}
for all $a>0$ and $0 \leq t_1<\dots<t_n$.
\end{proof}

\vspace{.5cm}

\subsection{Variance of the estimator}
\noindent Concerning the variance of estimator \eqref{eq:empirical_H}, we prove the following
\begin{prop}[\textit{Variance of $\hat{H}_0$}]\label{prop:varH}
    If process $Z_t$ has normally distributed increments decorrelated by random permutation, the variance of estimator \eqref{eq:distr_H-sssi3} is 
    \begin{equation} \label{eq:variance}
        Var(\hat{H}_0) \approx \frac{2\pi e}{(\ln a)^2}\left(\frac{1}{\sqrt{n}}+\frac{1}{\sqrt{m}}\right)^2.
    \end{equation}
\end{prop}
\begin{proof}
    To calculate the variance of $\hat{H}_0$ we can first evaluate the sensitivity of $\delta_{Z^{H_0}_{t}}(\Psi_H)$ to changes in $H$. This because, since the variance of $\Phi_{Z_{t,1}^{H_0}}$ equals $1$, changing $H$ only affects the variance of the scaled process $Z_{t,\overline{a}}^{H_0}$, which in turn affects the diameter by changing the point $\overline{a}^{H-H_0}x$. In general let us denote by $\Phi(x;v)$ the cumulative distribution function (CDF) of a Gaussian r.v. with mean $0$ and variance $v$. The problem becomes to analyze how the diameter
\begin{equation}
    D(v)=\sup_x |\Phi(x;1)-\Phi(x;v)|
\end{equation}
 behaves as a function of $v$.
 
 First, observe that the maximum absolute difference between the two CDFs theoretically occurs at points $\pm \sqrt{\frac{\ln v}{1-1/v}}$, i.e. at the two points of intersection between the corresponding densities. This means that for empirical CDFs the maximum absolute difference occurs somewhere around these points.
First, let us approximate the diameter for small deviations in $v$ from $1$. Set $v=1+\epsilon$, where $\epsilon$ is small and expand $\Phi(x;1+\epsilon)$ around $\epsilon=0$:
    \begin{equation}
        \Phi(x;1+\epsilon) \approx \Phi(x;1)+\epsilon \cdot \frac{\partial{\Phi}}{\partial v}(x;1).
    \end{equation}
    Since $\Phi$ is Gaussian, the derivative evaluated at $v=1$ is
\begin{equation}
    \frac{\partial{\Phi}}{\partial v}(x;1) = -\frac{xe^{-x^2/2}}{2\sqrt{2\pi}}=-\frac{x}{2}\phi(x),
\end{equation}
    where $\phi(x)$ denotes the standard normal density.
    Thus, for small $\epsilon$, 
    \begin{equation}
        \Phi(x;1+\epsilon)-\Phi(x;1) \approx -\frac{\epsilon}{2}x\phi(x)
    \end{equation}
    and therefore the absolute difference between CDFs is approximately linear in $\epsilon$ with coefficient $-\frac{1}{2}x\phi(x)$, i.e.
    \begin{equation}\label{eq:diamapprox}
        D(1+\epsilon) \approx \sup_x\left|-\frac{\epsilon}{2}x\phi(x)\right| = \frac{|\epsilon|}{2} \sup_x\left|x\phi(x)\right|.
    \end{equation}
    Function $x\phi(x)$ is maximized at $x=\pm 1$, where it attains $\phi(\pm1) = \pm \frac{1}{\sqrt{2\pi}}e^{-1/2}$. This gives the approximate diameter for small variance perturbations. Since in our case the variance perturbation is not $\epsilon$ but $a^{2(H_0-H)}-1$, setting $v(H)=a^{2(H_0-H)}$, for $H$ close to $H_0$, we can write:
    \begin{equation}
        v(H)=a^{2(H_0-H)}=e^{2(H_{0}-H) \ln a} \approx 1+ 2(H_0-H) \ln a .
    \end{equation}
    Therefore, the perturbation in variance is approximately $\epsilon = 2(H_0-H) \ln a$. Plugging this into \eqref{eq:diamapprox}, one has that the diameter is approximately linear in $|H-H_0|$ with slope $|\ln a|/\sqrt{2\pi e}$. This is the theoretical approximation for the diameter between two Gaussian distribution with variances $1$ and $a^{2(H_0-H)}$.\\
    Since we deal with empirical CDFs based on finite samples, the DKW inequality\footnote{The Dvoretzky–Kiefer–Wolfowitz–Massart inequality (DKWM)
    \begin{equation}\nonumber
        P\left(\sup_{x\in \mathbb{R}}|F_n(x)-F(x)|>\epsilon\right) \leq 2e^{-2n\epsilon^2} \quad \text{for any }\epsilon>0
    \end{equation}
    provides a bound on the worst case distance of the empirical distribution function $F_n(x)$ from its associated population distribution function $F(x)$. The inequality, which strengthens the Glivenko–Cantelli theorem by quantifying the rate of convergence as $n$ tends to infinity, implies that the empirical CDF converges uniformly to the true CDF at a rate $O_p(1/\sqrt{n})$. In the two sample-case the DKWM inequality holds for $m=n\geq 458$, see \citep{WeiDudley2012}.} applies, therefore the fluctuations around the theoretical value are of order $O_p\left(1/\sqrt{n}+1/\sqrt{m}\right)$. In fact, in the two-sample case, to bound the difference between the empirical and theoretical diameter, one can decompose the error using the triangle inequality:
    \begin{equation}
        \sup_x|F_n(x)-G_m(x)| \leq \underbrace{ \sup_x|F_n(x)-F(x)|}_\text{Error in $F_n$} + \underbrace{ \sup_x|F(x)-G(x)|}_\text{Theoretical diameter}+\underbrace{ \sup_x|G(x)-G_m(x)|}_\text{Error in $G_m$}. 
    \end{equation}
    At $H=H_0$, $\sup_x|F(x)-G(x)|=0$, therefore
    \begin{eqnarray}
        \sup_x|F_n(x)-G_m(x)| &\leq& \underbrace{ \sup_x|F_n(x)-F(x)|}_\text{Error in $F_n$} +\underbrace{ \sup_x|G(x)-G_m(x)|}_\text{Error in $G_m$}. \nonumber \\
        & & \quad \quad O_p\left(\frac{1}{\sqrt{n}}\right) \quad \quad \sep + \quad \quad O_p\left(\frac{1}{\sqrt{m}}\right)
    \end{eqnarray}
  Thus the total fluctuation is $O_p\left(\frac{1}{\sqrt{n}}+\frac{1}{\sqrt{m}}\right)$.\\
To minimize the empirical KS distance, we balance the bias term $D\left(a^{2(H_0-H)}\right)\approx |H_0-H||\ln a|/\sqrt{2\pi e}$ and the variance term $O_p\left(1/\sqrt{n}+1/\sqrt{m}\right)$. Setting the bias equal to the standard deviation:
    \begin{equation}
        |H_0-H|\frac{|\ln a|}{\sqrt{2\pi e}} \sim \frac{1}{\sqrt{n}}+\frac{1}{\sqrt{m}}.
    \end{equation}
It follows
    \begin{equation}
        \hat{H}_0-H_0 \sim \frac{\sqrt{2\pi e}}{|\ln a|}\left(\frac{1}{\sqrt{n}}+\frac{1}{\sqrt{m}}\right)Z,
    \end{equation}
where $Z \sim N(0,1)$. This entails 
that the asymptotic variance is 
\begin{equation} \label{eq:Varest}
    \sigma^2=\text{Var}(\hat{H}_0) \approx \frac{2 \pi e}{(\ln a)^2} \cdot \left(\frac{1}{\sqrt{n}}+\frac{1}{\sqrt{m}} \right)^2
\end{equation}
and the rate of convergence of $\hat{H}_0$ to $H_0$ is $O_p(1/\sqrt{n}+1/\sqrt{m})$.
\end{proof}

Thus, extracting two rescaled and randomized samples $Z_{1,\cdot}$ and $Z_{2,\cdot}$, of length $n$ and $m$ respectively and referred to time scales $\underline{a}$ and $\overline{a}$, with empirical CDFs $\Phi_{1,n}$ and $\Phi_{2,m}$, the empirical diameter is 
\begin{equation}\label{eq:empirical_delta}
    \hat{\delta}_{Z_t}\left(\Psi_H\right) = \sup\limits_{x\in\mathbb{R}}\left\vert \Phi_{1,n}(x) - \Phi_{2,m}(x)\right\vert = \sup\limits_{x\in\mathbb{R}}\left\vert \frac{1}{n}\sum\limits_{i=1}^n\mathbbm{1}_{\underline{a}^{-H}Z_{1,i}\leq x}-\frac{1}{m}\sum\limits_{i=1}^m\mathbbm{1}_{\overline{a}^{-H}Z_{2,i}\leq x}\right\vert
\end{equation}
and the estimated self-similarity parameter is $\hat{H}_0 = \argmin\limits_{H\in\left(0,1\right]}\,\,\hat{\delta}_{Z_t}\left(\Psi_H\right)$, provided that $\hat{\delta}_{X_t}\left(\Psi_H\right)$ can be considered negligible by comparison with the critical value of the KS test (\citep{AngeliniBianchi2025}).\\

\subsubsection{Variance reduction}\label{subsubsec:varRed}
In order to stabilize the estimates and reduce the variance we adopt the repeated subsampling with averaging, with random subsamples of length $L\ll m$. On each selection we compute the estimator $\hat{H}^{(k)}$ of the parameter $H$; by the above properties it is $\mathbb{E}[\hat{H}^{(k)}] = H$ and $\mathrm{Var}(\hat{H}^{(k)}) = \sigma^2$. Repeating the estimation $K$ times, we take as final estimator the repeated subsampling estimator
\begin{equation*}
    \overline{H} = \frac{1}{K}\sum_{k=1}^{K} \hat{H}^{(k)}.
\end{equation*}
Clearly,
\begin{eqnarray*}
    \mathrm{Var}(\overline{H}) 
&=& \mathrm{Var}\left(\frac{1}{K}\sum_{k=1}^K \hat{H}^{(k)}\right)
= \frac{1}{K^2}\,\mathrm{Var}\left(\sum_{k=1}^K \hat{H}^{(k)}\right) \nonumber \\
&=& \frac{1}{K^2}\,\left[\sum_{k=1}^K \mathrm{Var}(\hat{H}^{(k)})
  + 2\sum_{1\le i<j\le K} \mathrm{Cov}(\hat{H}^{(i)},\hat{H}^{(j)})\right] \nonumber \\
  &=& \frac{1}{K^2}\left[
  K\sigma^2+ 2\,\sum_{1\le i<j\le K} \mathrm{Cov}(\hat{H}^{(i)},\hat{H}^{(j)})\right]. \nonumber
\end{eqnarray*}
The usual Monte Carlo variance reduction is therefore obtained if $\mathrm{Cov}(\hat{H}^{(i)},\hat{H}^{(j)}) = 0$ for $i\neq j$. If one assume constant pairwise correlation, $\mathrm{Cov}(\hat{H}^{(i)},\hat{H}^{(j)}) = \rho\,\sigma^2$, then of course
\begin{equation*}
    \mathrm{Var}(\overline{H}) = \frac{\sigma^2}{K}\left[1+(K-1)\,\rho\right]
\end{equation*}
Therefore, because $\rho>0$, 
\begin{equation*}
    \frac{\sigma^2}{K} \le \mathrm{Var}(\overline{H}) \le \sigma^2,
\end{equation*}
with the actual value depending on how strongly the subsamples overlap.

In practice, we estimate $\rho$ empirically from the $K$ estimates and compute the sample variance:
\begin{equation*}
    s^2 = \frac{1}{K-1}\sum_{k=1}^K \big(\hat{H}^{(k)} - \overline{H}\big)^2.
\end{equation*}
Since $\mathbb{E}[s^2] = \sigma^2(1-\rho)$, an approximate estimator of $\rho$ is
\begin{equation*}
\hat\rho \approx 1 - \frac{s^2}{\sigma^2}.    
\end{equation*}
Plugging $\hat\rho$ in
\begin{equation}\label{eq:variance_reduced}
    \widehat{\mathrm{Var}}(\overline{H}) \approx \frac{\sigma^2}{K}\big[1 + (K-1)\hat\rho\big]=\sigma^2+s^2\left(\frac{1}{K}-1\right),
\end{equation}
which provides the data-driven estimate of how much variance reduction is actually achieved.\\

\begin{figure}[h!]
    \centering
    \includegraphics[width=1\linewidth]{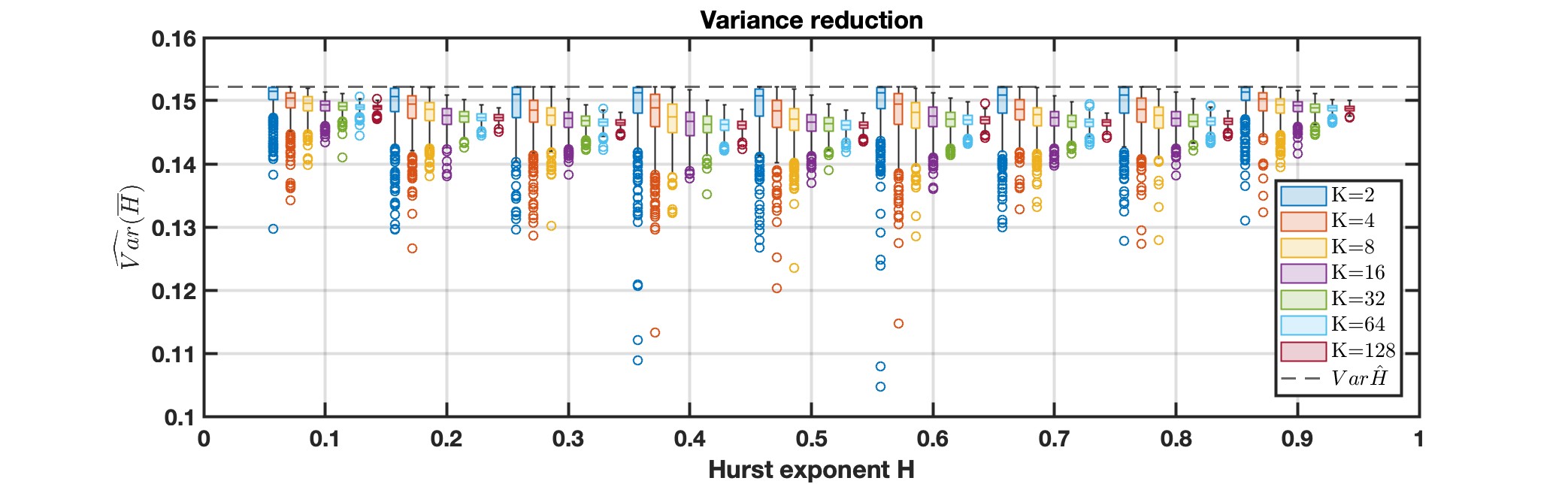}
    \caption{Variance reduction for a fBm with Hurst exponent $H\in[0.1,0.9]$. This variance reduction $\widehat{Var}(\overline{H})$ is the effect of $K$ repeated sampling of $T=50$ random sampling from a window with length $n=512$ and different Hurst exponent H. The distributions represent the variance reduction on equation \eqref{eq:variance_reduced} using 500 simulations. The black dashed line is the variance without repeated sampling $Var(\hat{H})$, i.e., the estimator variance described in proposition \ref{prop:varH}.} 
    \label{fig:variance_reduction}
\end{figure}

In order to evaluate the magnitude of this reduction, we performed 500 simulations for different Hurst exponent in the range $[0.1,0.9]$ and for a different number $K$ of repeated subsampling. In Figure \ref{fig:variance_reduction} we show the effect of $K$ repeated sumsampling of $T=50$ random sampling from a window with length $n=512$ on a fBm. The black dashed line represents the variance without repeated sampling (i.e., $K=1$) described in the proposition \ref{prop:varH}.
The reduction is linear up to $K=16$, after which it becomes non-linear, reaching a lower plateau. This is symptomatic of a correlation $\hat{\rho}$ between estimates $\hat{H}^{(k)}$ that prevails over the number of repeated sampling. Furthermore, it can be observed that for $H=0.5$, where there is less correlation between the estimates, the distributions are slightly lower than for $H\neq 0.5$.

\section{Optimization and Computational Efficiency of the Estimation Method}\label{Optimization}
The estimation of $H$ can be approached using the methodology described in Section \ref{Method}. A natural starting point is the Grid Search (GS) procedure, which discretizes the interval $(0,1]$ into an equally spaced grid, evaluates the objective function at each grid point, and selects the value of 
$H$ that minimizes the Kolmogorov–Smirnov statistic (see Algorithm \ref{alg:GS}). While GS is straightforward and robust, it rapidly becomes computationally intensive when a fine grid is required. In particular, the computational cost increases linearly with the number of grid points, rendering the method inefficient for large-scale experiments or repeated estimations.

\begin{algorithm}[H]
\caption{Grid\_Search.exe}\label{alg:GS}
\begin{algorithmic}
\footnotesize
\LComment{Grid Search (GS) method to estimate the $\hat{H}_0$ that minimizes the KS statistic $\hat{\delta}_{Z_t}\left(\Psi_H\right)$ using the methodology described in the Section \ref{Method}}
\State
\State Setting $\Delta H$ as the grid step, we define a uniform mesh $H=[H_{start},1]$ with $H_{start}=\Delta H$.
\State The set of timescales $\pazccal{A}$ is completely defined by $\underline{a}=1$ and $\overline{a}>1$.
\State We define the fBm parameters: $H_0$ is the true Hurst parameter, $N$ is the length process.
\State Finally, we define the subsequence length $T$ for the random method.
\State
\LComment{Simulation of the fractional Brownian motion}
\State Simulation of a fBm $X^{H_0}$ using the \textit{fbmwoodchan(N,$H_0$)}\footnotemark
\State Computation of two fGn: $Z_{\cdot,1}^{H_0}$ and $Z_{\cdot,\overline{a}}^{H_0}$
\State
\LComment{Random permutation of $T$ values from the process $Z$ with function \textit{randperm}($Z,T$) [see prop.\ref{prop:1}]}
\State Randomization and permutation: $\tilde{Z}^{H_0}_{\cdot,1} = randperm(Z^{H_0}_{\cdot,1},T)$ and $\tilde{Z}^{H_0}_{\cdot,\overline{a}} = randperm(Z^{H_0}_{\cdot,\overline{a}},T)$
\State
\LComment{Minimum search}
\State{Initialize $i$ to 0}
\ForAll{$H_i$ ranging from $\Delta H$ to $1$ with step size $\Delta H$}
    \State Increment $i$ by 1
    \LComment{Computation of the KS statistic using the function kstest2($\cdot,\cdot$)}
    \State $\hat{\delta}_{\tilde{Z}_\cdot}\left(\Psi_{H_i}\right)$ = kstest2$\left(\tilde{Z}^{H_0}_{\cdot,1},\;\;\overline{a}^{H_i}\tilde{Z}^{H_0}_{\cdot,\overline{a}}\right)$
\EndFor
\LComment{Computation of the minimum of the function $\hat{\delta}_{\tilde{Z}_\cdot}\left(\Psi_H\right)$ and the grid step $i^\star$ relative to $\hat{H}_0$}
\State $\left[\hat{\delta}_{\tilde{Z}_\cdot}\left(\Psi_{H_0}\right),\;\; i^\star\right] = \min\left(\hat{\delta}_{\tilde{Z}_\cdot}\left(\Psi_{H}\right)\right)$
\State $\hat{H}_0 = H_{i^\star}$
\end{algorithmic}
\end{algorithm}
\footnotetext{The \textit{fbmwoodchan()} is a function based on the Wood and Chan circulating matrix method \cite{WoodChan1994}. In our work we used the MATLAB function available in the FracLab Toolbox 2.02 (INRIA package).} To address these limitations, we evaluate several well-established derivative-free optimization methods: Genetic Algorithm (GA) \cite{GeneticAlgorithm}, Simulated Annealing (SA) \cite{kirkpatrick1983optimization}, Particle Swarm (PS) \cite{parsopoulos2002recent}, Direct Search (DS) \cite{hooke1961direct}, Brent's Method (BM) \cite{brent2002algorithms}, and Nelder-Mead (NM) \cite{nelder1965simplex}. 

\begin{figure}[H]
    \centering
    \includegraphics[width=1\linewidth]{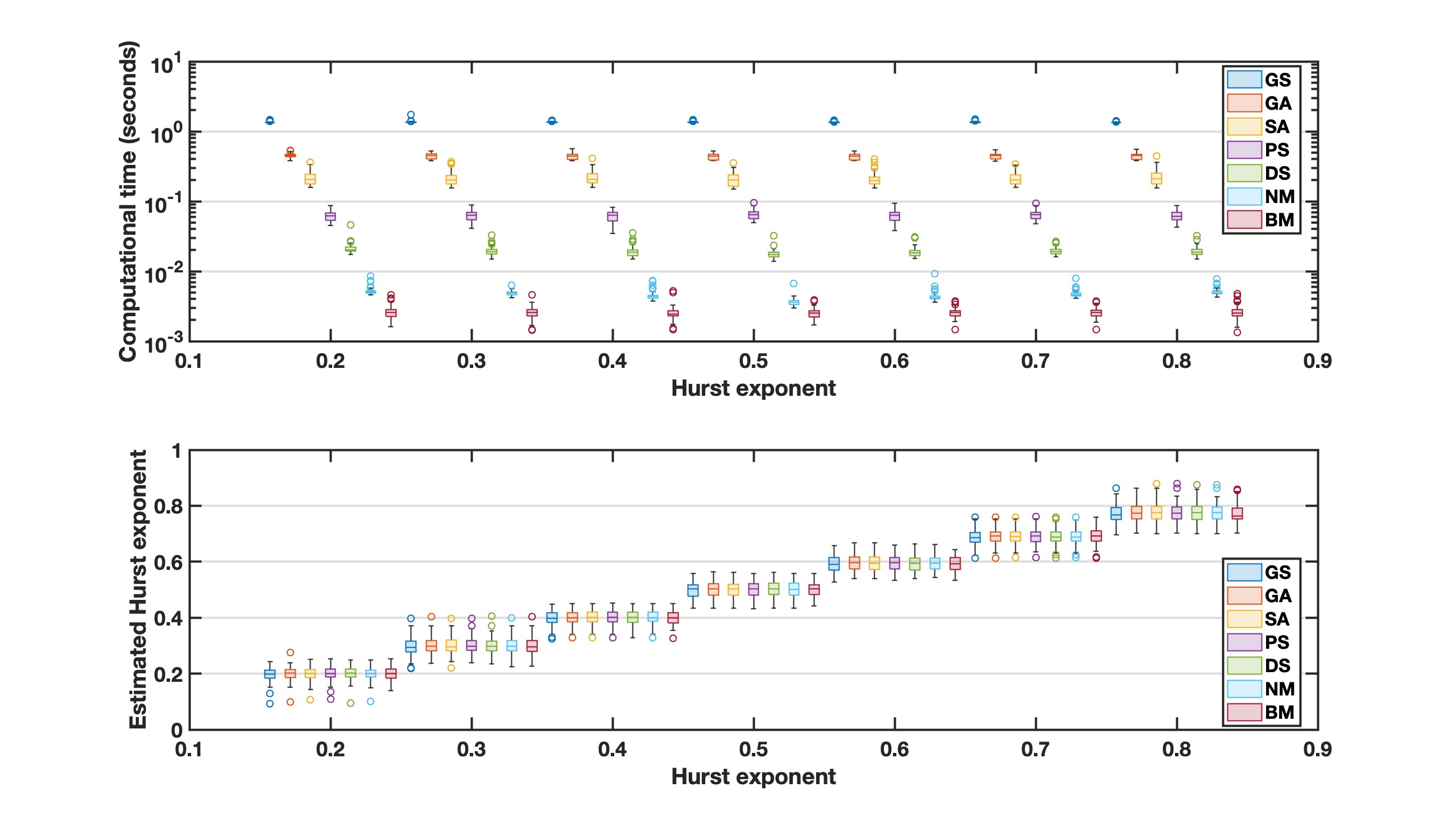}
    \caption{Top panel: Computational time (in seconds) distributions required by the seven optimization methods—Grid Search (GS), Genetic Algorithm (GA), Simulated Annealing (SA), Particle Swarm (PS), Direct Search (DS), Brent’s Method (BM), and Nelder–Mead (NM)—across different values of the Hurst parameter $H$. The GS displays markedly higher computational costs due to the discretization step, while derivative-free methods achieve substantially faster convergence. Bottom panel: Empirical distributions of the estimated Hurst parameter obtained with each optimization method, showing that despite differences in computational burden, all routines deliver comparable accuracy. Results are based on $100$ simulations.}
    \label{fig:optimization}
\end{figure}

In the implementation of the GS, we set the grid step to $\Delta H = 10^{-4}$, the upper scaling parameter $\overline{a} = 50$, the subsequence length $T = 500$, the process length $N = 2^{12}$. For the alternative optimization methods, instead of discretizing the parameter space, we imposed a tolerance level of $10^{-6}$. It is worth noting that, since GS operates on a fixed mesh of abscissae, no direct tolerance criterion was applied to the objective function $\delta\left(\Psi_H\right)$. Nevertheless, with the chosen parameters, the accuracy around the minimum value of $\delta\left(\Psi_H\right)$ was of the order $10^{-3}$. This level is acceptable because it does not alter the location of the estimated Hurst parameter, while guaranteeing numerical stability and keeping the GS computationally feasible for comparison with the other methods.

Figure \ref{fig:optimization} reports the computational times of the seven optimization methods across different values of the Hurst parameters $H\in[0.2,0.8]$ for $100$ simulations each. As expected, GS exhibits significantly higher computational costs with respect to the other algorithms. The derivative-free optimization methods achieve substantial improvements in efficiency while maintaining comparable accuracy. In particular, Brent's Method and Nelder-Mead stand out for their rapid convergence, making them attractive candidates for large-scale empirical applications where repeated estimations are required. By contrast, GS, although conceptually simple and robust, quickly becomes computationally impractical as the grid step is refined.

\section{Estimation pipeline}\label{Pipeline}
Before presenting the empirical results, we briefly summarize the generic estimation procedure adopted throughout the study, applicable to any observed time series (e.g., fBm-type Gaussian or Lévy processes).  
Given a time series $\{X_i\}_{i\in\llbracket 1,N\rrbracket}$, the Hurst exponent is estimated as follows:
\begin{enumerate}[leftmargin=*]\setlength\itemsep{.05cm}
    \item \textit{Sample segmentation.} The series is partitioned into $N\!-\!\nu$ overlapping windows of fixed length $\nu$. Each window is treated as a locally stationary realization of the underlying process.
    \item \textit{Multi-scale increments.} Within each window, increments are computed at two time scales, $\underline{a}=1$ and $\overline{a}>1$.
    \item \textit{Random permutation.} To mitigate temporal dependence, $T\ll\nu$ increments are randomly selected according to the permutation scheme of Section~\ref{sec:RandomPermutation}.
    \item \textit{Kolmogorov-Smirnov distance comparison.} For each candidate $H\in(0,1]$, the empirical Kolmogorov--Smirnov distance $\hat{\delta}$ between permuted rescaled increment distributions is evaluated.
    \item \textit{Parameter estimation.} $\hat{H}$ is obtained by minimizing the Kolmogorov--Smirnov distance (see~\eqref{eq:empirical_H}), using a derivative-free optimization algorithm. $\text{Var}(\hat{H})$ is computed as in equation~\eqref{eq:variance}.
    \item \textit{Variance reduction.} Steps 3--5 are repeated $K$ times and averaged, yielding $\overline{H}$ with reduced variance $\widehat{\Var}(\overline H)$ as shown in equation \eqref{eq:variance_reduced}.
    \item \textit{Iteration across windows.} The procedure is repeated for each rolling window, producing a sequence of local  estimates suitable for subsequent statistical analysis.
\end{enumerate}

\subsection{Time-variation filtering of the estimated Hurst parameter}
To distinguish genuine temporal variations from estimation noise, we use the following statistical filtering procedure (see e.g. \citep{Harvey1990}, p.125ff).  
Let $\{\overline{H}_t\}_{t=[1,T]}$ denote the averaged rolling-window estimates and $\{\sigma_t^2\}_{t=[1,T]}$ their associated reduced variances derived from Proposition \ref{prop:varH} after the variance reduction described in Section \ref{subsubsec:varRed}. We model $\overline{H}_t$ as a noisy observation of the latent process $\{H_t\}$ through the linear Gaussian state--space system
\begin{equation}
\begin{aligned}
    \overline{H}_t &= H_t + \varepsilon_t, & \varepsilon_t &\sim \mathcal N(0,\sigma_t^2),\\
    H_t &= H_{t-1} + \epsilon_t, & \epsilon_t &\sim \mathcal N(0,q),
\end{aligned}
\end{equation}
where $q\geq 0$ controls the degree of temporal variability of the latent process. The null hypothesis of parameter constancy corresponds to $q=0$ ($H_t = H$ for all $t$), while $q>0$ allows for gradual time variation via a random walk.

The model is estimated by maximum likelihood using the Kalman filter, which yields the exact Gaussian likelihood of $\{\overline{H}_t\}$ and optimal filtering and smoothing estimates of $\{H_t\}$. Time variation is assessed via the likelihood ratio test
\begin{equation}\label{eq:Kalman_test}
    H_0:q=0 \quad \text{versus} \quad H_1:q>0.
\end{equation}
Because the null lies on the boundary of the parameter space, the test statistic follows a $50$--$50$ mixture of a point mass at zero and a $\chi^2_1$ distribution, providing a conservative test for constancy.

\paragraph{Simulation evidence}
The state–space formulation explicitly disentangles genuine time variation in the Hurst parameter from fluctuations induced by estimation noise, by exploiting the known heteroskedastic variances of the estimator.

\begin{figure}[h!]
    \centering
    \includegraphics[width=0.9\linewidth]{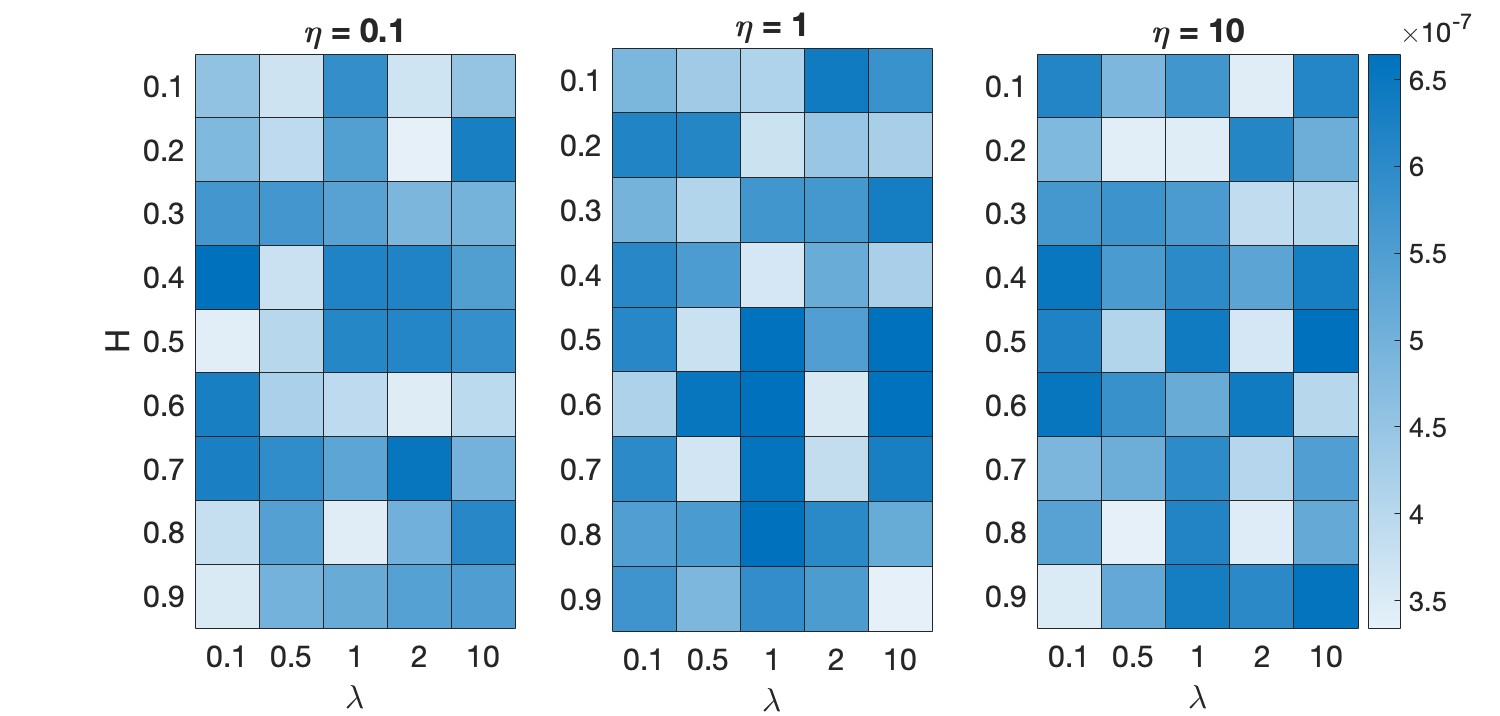}
    \caption{Heatmaps of the estimated variance $q$ from the state–space test applied to a fractional Ornstein–Uhlenbeck process as function of three parameters: Hurst exponent $H\in\{0.1,\ldots,0.9\}$ with $\Delta H=0.1$, the diffusion parameter $\eta = \{0.1,1,10\}$ and the mean-reversion coefficient $\lambda=\{0.1,0.5,1,2,10\}$.}
    \label{fig:Kalman_test}
\end{figure}

The stability of the test \eqref{eq:Kalman_test} is assessed by performing the pipeline estimation described above on a fOU for different parameters: The Hurst exponent $H\in\{0.1,\ldots,0.9\}$ with $\Delta H = 0.1$, the diffusion parameter $\eta = \{0.1,1,10\}$ and the mean-reversion term $\lambda = \{0.1, 0.5,1,2,10\}$. When $\lambda>1$, the process is strongly mean-reverting and rapidly stationary. Conversely, for $\lambda\ll1$, the dynamics approach those of a fractional Brownian motion, as the mean-reversion effect becomes negligible.\\

In Figure \ref{fig:Kalman_test}, three heatmaps are reproduced with the estimation for the variance $q$ of the latent process for any fOU's parameter: in each case, the estimation is on the order of $10^{-7}$. Even if this is not sufficient to state that the process $H_t$ is constant (indeed, the latent process can have minor changes guided by a small variance), the p-value of each estimate equals 0.5, indicating that the null hypothesis cannot be rejected, i.e. for each set of parameters, the fOU has a constant Hurst exponent, as expected.

\section{Implementation and data analysis} \label{Implementation}
Estimator \eqref{eq:empirical_H} was employed to obtain static estimate of the parameter $H$ using a four-market-year window of $\nu=1008$ observations of:
\begin{itemize}
    \item the logarithm of the CBOE VIX Index, spanning from January 3, 2000, to September 17, 2025, for a total of 6,493 observations;
    \item the logarithm of 5-minute realized volatility (RV5) of the S\&P 500 (SPX), from January 3, 2000 to June 27, 2018, totaling 4,713 observations.
\end{itemize}
The temporal endpoint of the analysis for the S\&P 500 realized volatility series (27 June 2018) is determined by data availability in the Oxford-Man Institute Realized Library. This database, while a standard reference in the academic literature for high-frequency financial data, is not systematically updated beyond this date.\\

\begin{figure}[h!]
    \centering
    \includegraphics[width=1\linewidth]{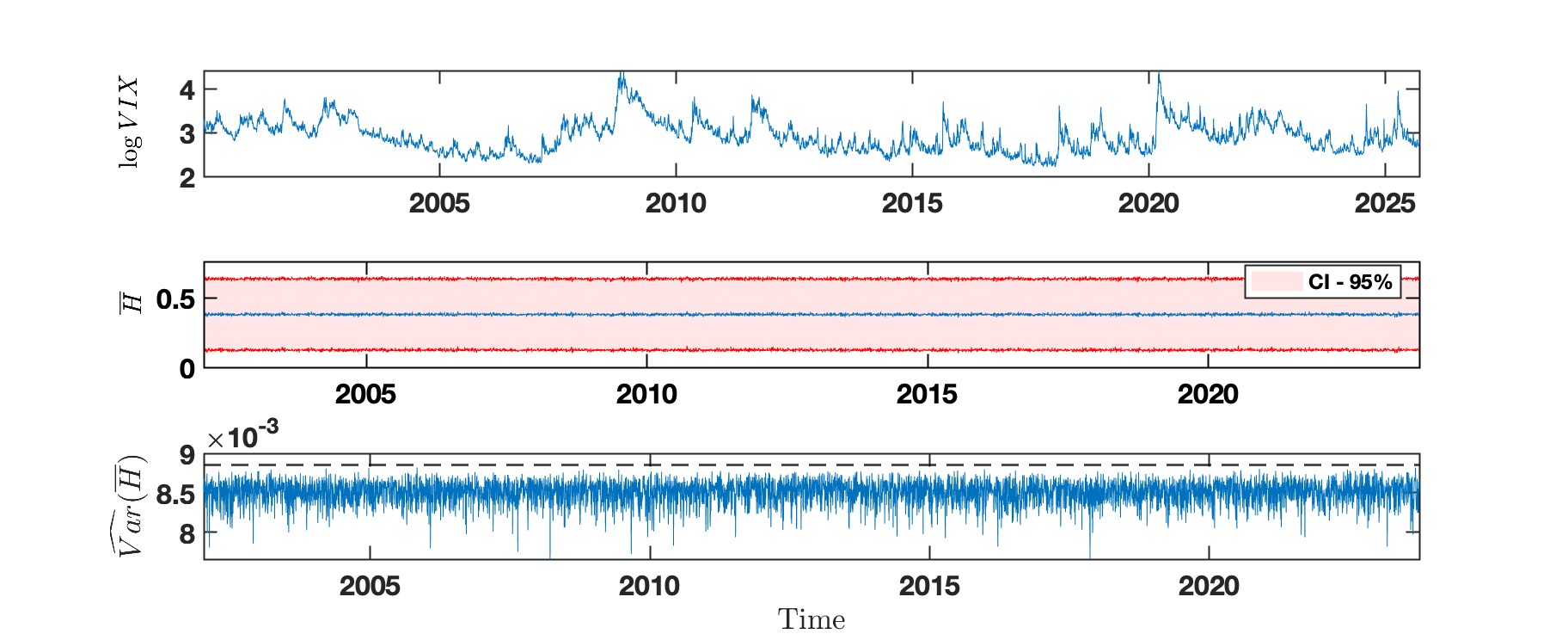}
    \caption{\textit{Randomized Kolmogorov-Smirnov analysis for VIX:} top panel shows the logarithm of the VIX process, the mid panel shows the Hurst exponent $\overline{H}_{VIX}$ estimated with the randomized KS test with its 95\% confidence interval and the bottom panel represents the variance reduced after $K=16$ repeated subsampling. Black dashed line represents the variance without resampling.}
    \label{fig:VIX_analysis}
\end{figure}

\begin{figure}[h!]
    \centering
    \includegraphics[width=1\linewidth]{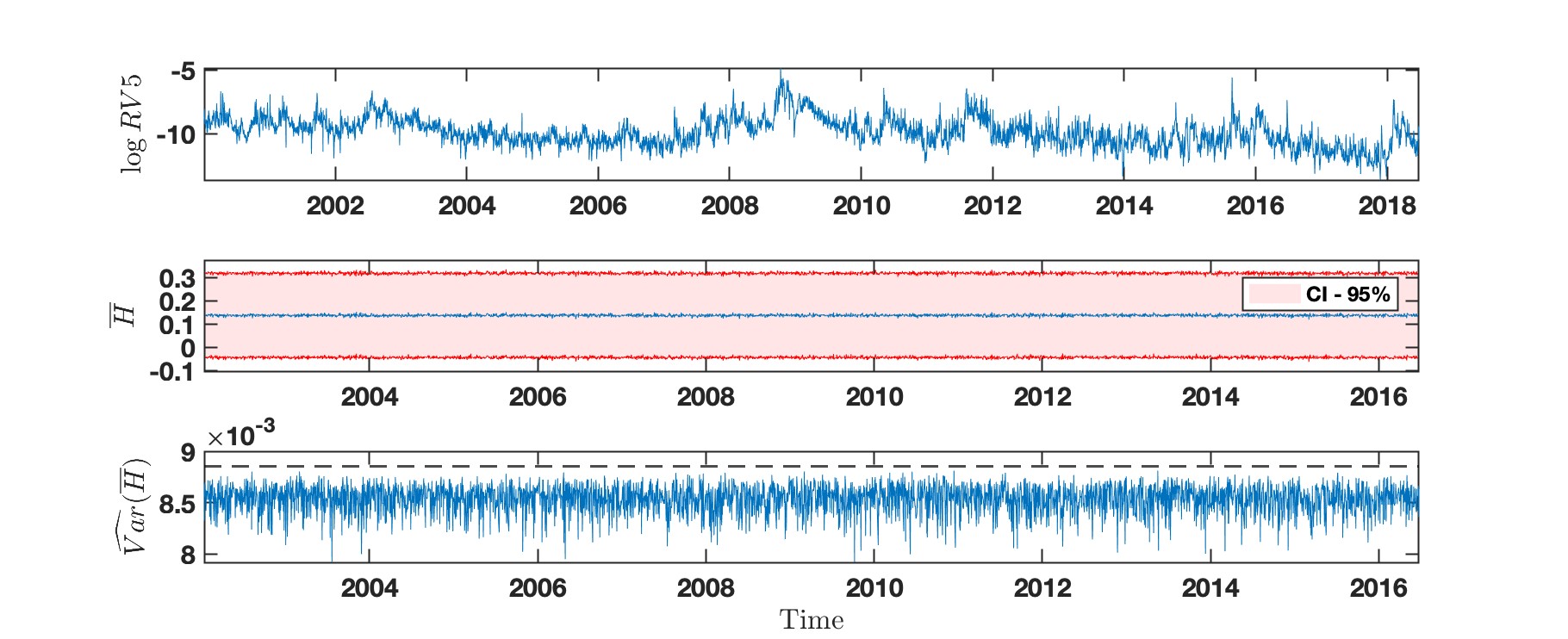}
    \caption{\textit{Randomized Kolmogorov-Smirnov analysis for Realized volatility:} top panel shows the logarithm of the RV5 process, the mid panel shows the Hurst exponent $\overline{H}_{RV5}$ estimated with the randomized KS test with its 95\% confidence interval and the bottom panel represents the variance reduced after $K=16$ repeated subsampling. Black dashed line represents the variance without resampling.}
    \label{fig:RV5_analysis}
\end{figure}

Estimator \eqref{eq:empirical_delta} was applied to the logarithm of VIX and realized volatility series with parameters $\underline{a}=1$, $\overline{a}=50$ and a subsequence length $T=508$, two-years market horizon, which entails a 95\%-confidence interval of $\hat{H}_0 \pm 0.1837$. Using $K=16$ repeated random sampling for any window, the confidence interval reduces as function of the empirical autocorrelation $\hat{\rho}$ between estimations. In Figures \ref{fig:VIX_analysis} and \ref{fig:RV5_analysis}, for the VIX and realized volatility respectively, we show the logarithm of the process (top panels), the Hurst estimator $\overline{H}_0$ with its 95\% confidence interval (middle panels), and the variance-reduced series $\widehat{Var}(\overline{H}_0)$ (bottom panels).


\begin{table}[h!]
    \centering
    \caption{The table reports in the first column the sample mean. Second column report the mean of the variance reduced. Third and fourth columns report, respectively, variance of latent process and p-value of the state-space test.}
    \begin{tabular}{|c|cccc|}
    \hline
        Process & $\overline{H}_{\text{avg}}$ & $\widehat{Var}(\overline{H})_{\text{avg}}$& $\hat{q}$ & p-value\\
        \hline
        \hline
        VIX & 0.3793 & 0.0085 & $4.8665\times 10^{-7}$ & 0.5000\\
        RV5 & 0.1379 & 0.0085 & $3.9064\times 10^{-7}$ & 0.5000\\
        \hline
    \end{tabular}
    \label{tab:Kalman_Test_Analysis}
\end{table}
Table \ref{tab:Kalman_Test_Analysis} summarizes the main statistics of the Hurst exponent estimates obtained from the randomized Kolmogorov–Smirnov procedure, together with the results of the state–space constancy test. Both volatility measures exhibit Hurst exponents significantly below $1/2$, confirming the presence of pronounced roughness in volatility dynamics. Consistently with previous empirical evidence, the estimated Hurst parameter for the VIX is substantially higher than that of realized volatility, indicating that implied volatility is smoother than its realized counterpart. The average variance of the estimator is comparable across the two series, reflecting the stability induced by the repeated subsampling and variance-reduction scheme. Finally, the estimated variance parameter of the latent state process is extremely small in both cases, and the associated p-values indicate that the null hypothesis of a constant Hurst exponent cannot be rejected. This suggests that the observed fluctuations in the rolling estimates are primarily attributable to estimation noise rather than genuine temporal variation in volatility roughness.

\section{Conclusions and Future Developments}\label{Conclusion}
This paper has introduced a distribution-based estimator for the self-similarity parameter of log-volatility processes, grounded in the Kolmogorov--Smirnov framework. By comparing entire distributions of rescaled increments rather than relying on moment-based scaling relations, the proposed methodology captures richer structural information and is designed to be more robust to the nonlinear biases that can affect traditional estimators.

A central methodological contribution concerns the treatment of the strong temporal dependence inherent in volatility data. We introduce a random permutation procedure which, for sufficiently large permutation lengths, effectively decorrelates the series while preserving marginal distributions (Proposition~\ref{prop:commute}). This device enables the valid application of the KS statistic to dependent financial time series and supports statistical inference. We further derive the asymptotic variance of the estimator (Proposition~\ref{prop:varH}), establishing a convergence rate of $O_p(1/\sqrt{n}+1/\sqrt{m})$ and providing a basis for confidence interval construction. Estimation stability is enhanced through a repeated subsampling and variance-reduction scheme.

From a computational perspective, we show that derivative-free optimization methods -- most notably Brent’s method and the Nelder--Mead simplex -- achieve substantial efficiency gains over naive grid search while delivering comparable accuracy. This feature is essential for large-scale empirical studies and rolling-window applications.

Empirically, we apply the estimator to the VIX index and to five-minute realized volatility of the S\&P~500. Consistent with previous results \citep{Livieri02092018,floch2022}, we find that the VIX displays a higher Hurst exponent ($\hat{H}_{\mathrm{VIX}} \approx 0.40$) than realized volatility ($\hat{H}_{\mathrm{RV}} \approx 0.14$), indicating that implied volatility is substantially smoother. At the same time, both measures exhibit Hurst exponents well below $1/2$, reinforcing the empirical relevance of rough volatility dynamics. A state--space filtering analysis further suggests that the observed fluctuations in rolling estimates are primarily driven by estimation noise rather than genuine time variation in the roughness parameter.

These findings highlight a fundamental open issue in fractional volatility modeling: the intrinsic coupling between local roughness and long-range dependence encoded in the Hurst parameter. Disentangling these two features remains a key challenge and motivates the development of models capable of separating short-scale regularity from long-memory effects \citep{Bennedsenetal2021,Eliazar2024}. Future research directions include extending the KS-based framework to settings that retain partial dependence information, integrating alternative volatility proxies and noise-corrected estimators, and adapting the methodology to multifractional or regime-switching environments in which local regularity may evolve over time.

\vspace{30pt}
\noindent \textbf{Acknowledgements \& Funding}. This research was supported by Sapienza University of Rome under Grant No. RM120172B346C021.

\begin{footnotesize}
    \bibliographystyle{plain}
\bibliography{Bibliography} 

\begin{thebibliography}{10}

\bibitem{abry2009wavelet}
P.~Abry, P.~Gonçalves, and D.~Veitch.
\newblock Wavelet-based estimation of the self-similarity parameter of long-range dependent signals.
\newblock {\em Statistical Methodology}, 6(1):33--52, 2009.

\bibitem{Alòsetal2007}
E.~Alòs, J.~A. Le{\'o}n, and J.~Vives.
\newblock On the short-time behavior of the implied volatility for jump-diffusion models with stochastic volatility.
\newblock {\em Finance and Stochastics}, 11(4):571--589, 2007.

\bibitem{Andersenetal2003}
T.~G. Andersen, T.~Bollerslev, F.~X. Diebold, and P.~Labys.
\newblock Modeling and forecasting realized volatility.
\newblock {\em Econometrica}, 71(2):579--625, 2003.

\bibitem{AngeliniBianchi2023}
D.~Angelini and S.~Bianchi.
\newblock Nonlinear biases in the roughness of a {f}ractional {s}tochastic {r}egularity {m}odel.
\newblock {\em Chaos, Solitons \& Fractals}, 172:113550, 2023.

\bibitem{AngeliniBianchi2025}
D.~Angelini and S.~Bianchi.
\newblock Kolmogorov--{S}mirnov estimation of self-similarity in long-range dependent fractional processes.
\newblock {\em Physica D: Nonlinear Phenomena}, 476:134697, 2025.

\bibitem{Baillie1996}
R.~T. Baillie.
\newblock Long memory processes and fractional integration in econometrics.
\newblock {\em Journal of Econometrics}, 73(1):5--59, 1996.

\bibitem{Baillieetal2019}
R.~T. Baillie, F.~Calonaci, D.~Cho, and S.~Rho.
\newblock Long memory, realized volatility and heterogeneous autoregressive models.
\newblock {\em Journal of Time Series Analysis}, 40(4):609--628, 2019.

\bibitem{bayer2021deep}
C.~Bayer, B.~Horvath, and A.~Muguruza.
\newblock Deep calibration of rough stochastic volatility models.
\newblock {\em Quantitative Finance}, 21(1):25--41, 2021.

\bibitem{Bennedsenetal2021}
M.~Bennedsen, A.~Lunde, and M.~S. Pakkanen.
\newblock Decoupling the short- and long-term behavior of stochastic volatility.
\newblock {\em Journal of Financial Econometrics}, 20(5):961--1006, 2022.

\bibitem{bennedsen2021learning}
M.~Bennedsen, M.S. Pakkanen, and A.~Lunde.
\newblock Learning the roughness of stochastic volatility from option prices.
\newblock {\em Journal of Financial Econometrics}, 19(3):403--431, 2021.

\bibitem{Bianchi2004}
S.~Bianchi.
\newblock A new distribution-based test of self-similarity.
\newblock {\em Fractals}, 12(03):331--346, 2004.

\bibitem{bianchi2025roughness}
S.~Bianchi and D.~Angelini.
\newblock Roughness in vix index and in realized volatility: Rolling window estimation by randomized {K}olmogorov-{S}mirnov distribution.
\newblock {\em Springer Books}, pages 61--73, 2025.

\bibitem{BolkoChristensenPakkanenVeliyeva2022}
E.~Bolko, K.~Christensen, M.S. Pakkanen, and B.~Veliyeva.
\newblock A {GMM} approach to estimate the roughness of stochastic volatility.
\newblock {\em Journal of Econometrics}, Available online 15 September 2022, 2022.

\bibitem{brent2002algorithms}
R.~Brent.
\newblock {\em Algorithms for {M}inimization {W}ithout {D}erivatives}.
\newblock Prentice-Hall, 1973.

\bibitem{CheriditoKawaguchiMaegima2003}
P.~Cheridito, H.~Kawaguchi, and M.~Maejima.
\newblock Fractional {O}rnstein-{U}hlenbeck processes.
\newblock {\em Electronic Journal of Probability}, 8:1--14, 2003.

\bibitem{Coeurjolly2000}
J.-F. Coeurjolly.
\newblock Simulation and identification of the fractional {B}rownian motion: {A} bibliographical and comparative study.
\newblock {\em Journal of Statistical Software}, 5:1--53, 2000.

\bibitem{ComteRenault1998}
F.~Comte and E.~Renault.
\newblock Long memory in continuous-time stochastic volatility models.
\newblock {\em Mathematical Finance}, 8(4):291--323, 1998.

\bibitem{Cont2001}
R.~Cont.
\newblock Empirical properties of asset returns: stylized facts and statistical issues.
\newblock {\em Quantitative Finance}, 1(2):223--236, 2001.

\bibitem{ContDas2022}
R.~Cont and P.~Das.
\newblock Quadratic variation along refining partitions: constructions and examples.
\newblock {\em J. Math. Anal. Appl.}, 512(2):126173, 2022.

\bibitem{Ding1993}
Z.~Ding, C.~W.~J. Granger, and R.~F. Engle.
\newblock A long memory property of stock market returns and a new model.
\newblock {\em Journal of Empirical Finance}, 1(1):83--106, 1993.

\bibitem{Eliazar2024}
I.~Eliazar.
\newblock Power {B}rownian motion.
\newblock {\em Journal of Physics A: Mathematical and Theoretical}, 57(3):03LT01, dec 2023.

\bibitem{Fukasawa2019}
M.~Fukasawa, T.~Takabatake, and R.~Westphal.
\newblock Is volatility rough?
\newblock {\em arXiv preprint arXiv:1905.04852}, 2019.

\bibitem{Fukasawaetal2022}
M.~Fukasawa, T.~Takabatake, and R.~Westphal.
\newblock Consistent estimation for fractional stochastic volatility model under high-frequency asymptotics.
\newblock {\em Mathematical Finance}, 32(4):1086--1132, 2022.

\bibitem{GatheralJaissonRosenbaum2018}
J.~Gatheral, T.~Jaisson, and M.~Rosenbaum.
\newblock Volatility is rough.
\newblock {\em Quantitative Finance}, 18(6):933--949, 2018.

\bibitem{GeneticAlgorithm}
D.~E. Goldberg.
\newblock {\em Genetic Algorithms in Search, Optimization, and Machine Learning}.
\newblock Addison-Wesley, 1989.

\bibitem{Harvey1990}
A.C. Harvey.
\newblock {\em Forecasting, Structural Time Series Models and the Kalman Filter}, page 125.
\newblock Cambridge {U}niversity {P}ress, 1990.

\bibitem{hooke1961direct}
R.~Hooke and T.~A. Jeeves.
\newblock ``{D}irect search'' solution of numerical and statistical problems.
\newblock {\em Journal of the ACM (JACM)}, 8(2):212--229, 1961.

\bibitem{kirkpatrick1983optimization}
S.~Kirkpatrick, C.~D. Gelatt~Jr, and M.~P. Vecchi.
\newblock Optimization by simulated annealing.
\newblock {\em Science}, 220(4598):671--680, 1983.

\bibitem{LacazeRoviras2002}
B.~Lacaze and D.~Roviras.
\newblock Effect of random permutations applied to random sequences and related applications.
\newblock {\em Signal Processing}, 82(6):821--831, 2002.

\bibitem{floch2022}
F.~Le~Floc'h.
\newblock Roughness of the implied volatility, arxiv, 2207.04930, 2022.

\bibitem{Livieri02092018}
G.~Livieri, S.~Mouti, A.~Pallavicini, and M.~Rosenbaum.
\newblock Rough volatility: Evidence from option prices.
\newblock {\em IISE Transactions}, 50(9):767--776, 2018.

\bibitem{nelder1965simplex}
J.~A. Nelder and R.~Mead.
\newblock A simplex method for function minimization.
\newblock {\em The Computer Journal}, 7(4):308--313, 1965.

\bibitem{parsopoulos2002recent}
K.~E. Parsopoulos and M.~N. Vrahatis.
\newblock Recent approaches to global optimization problems through particle swarm optimization.
\newblock {\em Natural Computing}, 1(2):235--306, 2002.

\bibitem{PoonandGranger2003}
S.-H. Poon and C.~W. Granger.
\newblock Forecasting volatility in financial markets: A review.
\newblock {\em Journal of Economic Literature}, 41(2):478--539, 2003.

\bibitem{rosenbaum2019estimation}
M.~Rosenbaum and V.~Morel.
\newblock Estimation of roughness parameters: {A}pplication to rough volatility modeling.
\newblock {\em Finance and Stochastics}, 23(2):501--533, 2019.

\bibitem{Sato1999}
Ken-Iti Sato.
\newblock {\em L{é}vy {P}rocesses and {I}nfinitely {D}ivisible {D}istributions}.
\newblock {C}ambridge {U}niversity {P}ress, 1999.

\bibitem{SanchezGraneroetal2012}
M.~J. Sánchez-Granero, M.~Fernández-Martínez, and J.~E. Trinidad-Segovia.
\newblock Introducing fractal dimension algorithms to calculate the {H}urst exponent of financial time series.
\newblock {\em The European Physical Journal B}, 85(86):1--13, 2012.

\bibitem{Trinidadetal2012}
J.~E. {Trinidad-Segovia}, M.~Fernández-Martínez, and M.~A. Sánchez-Granero.
\newblock A note on geometric method-based procedures to calculate the {H}urst exponent.
\newblock {\em Physica A: Statistical Mechanics and its Applications}, 391(6):2209--2214, 2012.

\bibitem{WeiDudley2012}
F.~Wei and R.~M. Dudley.
\newblock Two-sample {D}voretzky–{K}iefer–{W}olfowitz inequalities.
\newblock {\em Statistics \& {P}robability {L}etters}, 82(3):636--644, 2012.

\bibitem{WoodChan1994}
A.~T.~A. Wood and G.~Chan.
\newblock Simulation of stationary {G}aussian processes in [0, 1]d.
\newblock {\em Journal of Computational and Graphical Statistics}, 3(4):409--432, 1994.

\end{thebibliography}
\end{footnotesize}
\end{document}